\documentclass[prb,superscriptaddress,twocolumn,showpacs,preprintnumbers,amsmath,floatfix]{revtex4}
\usepackage{amssymb}
\usepackage{graphicx}

\begin{document}

\title{Possible Way to Make MgFeGe a New Fe-based Superconductor}
\author{Ming-Cui Ding}
\affiliation{Shanghai Key Laboratory of Special Artificial Microstructure Materials and Technology, \\
School of Physics Science and engineering, Tongji University, Shanghai 200092, P.R. China}
\author{Yu-Zhong Zhang}
\email[Corresponding author. ]{Email: yzzhang@tongji.edu.cn}
\affiliation{Shanghai Key Laboratory of Special Artificial Microstructure Materials and Technology, \\
School of Physics Science and engineering, Tongji University, Shanghai 200092, P.R. China}
\date{\today }

\begin{abstract}
\centerline{Abstract}

We propose that the contrasting low-temperature behaviors observed experimentally among isostructural and isoelectronic materials, like non-superconducting and nonmagnetic MgFeGe, magnetically ordered NaFeAs, and superconducting LiFeAs, can be well understood from itinerant weak coupling limit. We find that stronger $(\pi,\pi)$ instability appearing in the d$_{x^2-y^2}$ orbital of NaFeAs is responsible for the occurrence of weak magnetism while weaker but still prominent $(\pi,\pi)$ instability in LiFeAs leads to a superconducting state. In contrast, multiple competing instabilities coexisting in orbital-resolved momentum-dependent susceptibilities, serving as magnetic frustrations from itinerant electrons, may account for the nonmagnetic state in MgFeGe, while poorer Fermi surface nesting leads to a non-superconducting state. Based on above findings, we predict a possible way to make MgFeGe a new Fe-based superconductor.

\end{abstract}


\maketitle

\section{Introduction\label{Introduction}}

Intensive debates persist since the discovery of High-T$_c$ iron-based superconductors on how to understand the origin of magnetism and superconductivity, either from strong coupling localized limit or from weak coupling itinerant limit~\cite{Yildirim,Si,Ma,Schmidt,Ducatman,SinghDu,Mazinlafeaso,Zhang122,ZhangValenti,KnolleEremin,Ferber,PRX,scalapino,Pickett,Paglione,Wanglee,DaiHu,Dingrev}. Among various successfully synthesized iron-based compounds, several of them are presumed to be the  counterexamples against the itinerant scenario, such as potassium doped iron selenides~\cite{guochen,Zhaojun} and iron tellurides~\cite{baowei,yxia}.  However, on one hand, it is still unresolved experimentally which is the parent compound for superconductivity in potassium doped iron selenides due to the coexistence of different types of iron vacancy order and superlattices~\cite{kfe2se24,WeiLikfese,WangLi,Ricci,Caiwang,Daggoto}. Therefore it remains unknown if the itinerant scenario is really inapplicable to the potassium doped iron selenides. On the other hand, it has been shown from a density functional theory (DFT) calculation that the unique antiferromagnetism with double stripe observed in iron tellurides can be well interpreted from itinerant picture as long as excess interstitial irons are properly taken into account~\cite{zhangfete}.

MgFeGe, a new candidate of 111 family of iron-based superconductors in addition to LiFeAs and NaFeAs, now becomes another possible counterexample against the itinerant scenario. From experiments, it is found that MgFeGe is nonmagnetic and non-superconducting down to 2~K~\cite{hosonomgfege}, in contrast to LiFeAs which is a good superconductor with T$_c$ = 18~K~\cite{Tapplifeas} and to NaFeAs which shows a magnetically driven structural phase transition above the superconducting transition~\cite{dainafeas}. However, surprisingly, not only the lattice structures but also the electronic structures like band dispersions and Fermi surfaces of MgFeGe are all similiar to those of LiFeAs~\cite{hosonomgfege}. Furthermore, a quantitative calculation on momentum dependent Pauli susceptibility within constant matrix elements approximation showed that the $(\pi,\pi)$ instability which is responsible for stripe-type antiferromagnetic order or superconductivity is even stronger in MgFeGe than in LiFeAs~\cite{rheemgfege}, indicating that from itinerant point of view MgFeGe is more likely a superconductor or an antiferromagnet. This is inconsistent with the experimental findings. As a result, the applicability of weak coupling approach to the origin of magnetism or superconductivity in iron-based compounds is again questionable.

Recently, Jeschke, {\it et al.}~\cite{Jeschke2013} emphasized the importance of looking at magnetism in MgFeGe. Based on a DFT calculation where spin polarized generalized gradient approximation (GGA) is used, it was pointed out that the ground state of MgFeGe is ferromagnetically ordered if the experimental lattice structure is used, indicating that strong ferromagnetic fluctuations might be the reason for non-superconducting and nonmagnetic state. However, the result is inconsistent with existing ones. First, it was previously reported by applying similar calculations that the ground state of MgFeGe is of stripe-type antiferromagnetic order if optimized lattice structure is used~\cite{hosonomgfege}. Second, the calculated Pauli susceptibility doesn't show any instabilities at wave vector of $(0,0)$~\cite{rheemgfege}, indicating that there is no strong ferromagnetic fluctuation in the paramagnetic state.

Furthermore, we apply both GGA and local density approximation (LDA) to the DFT calculations~\cite{ourresult} and find that the nature of the ground state is strongly dependent on the functionals one chooses (see Table~\ref{Tab:one} and discussion below). As is well known that, while GGA usually underestimates the binding energy, LDA does in the opposite way. Thus, the contradiction between the results from GGA and LDA, together with the above mentioned conflicts, strongly imply that a new theory, which can eliminate all the inconsistencies, has to be established to understand the mechanism for the different behaviors among these isostructural and isoelectronic compounds.

In this paper, we will show that the differences among MgFeGe, LiFeAs and NaFeAs can be well explained from weak coupling limit irrespective of the functionals one uses if intraorbital contributions to the particle-hole scattering are individually considered. The presence or the absence of $(\pi,\pi)$ instability, which is widely accepted to be responsible for appearance and disappearance of the magnetism or superconductivity, in the d$_{x^2-y^2}$ orbital plays a dominant role in determining the low-temperature behaviors of these three compounds. While NaFeAs and LiFeAs show stronger and weaker, but both prominent, $(\pi,\pi)$ instabilities in d$_{x^2-y^2}$ orbital, respectively, MgFeGe exhibits multiple competing instabilities in the momentum space. As a consequence, NaFeAs is magnetically ordered at low temperature while MgFeGe is nonmagnetic. Moreover, the poorer Fermi surface nesting in MgFeGe than in superconducting LiFeAs leads to a non-superconducting state in MgFeGe. Finally, we predict that electron doping, such as substitution of Mg by Al or La, will be an efficient way to make MgFeGe a new iron-based superconductor.

\section{Results\label{Results}}

\subsection{Energies of various spin configurations\label{magnetism}}
\begin{table}
  \caption{Energies of various spin configurations
    with respect to the nonmagnetic solution for MgFeGe within LDA and GGA, in meV/Fe. Spins are arranged ferromagnetically along $c$ axis in the first four states while antiferromagnetically in the rest four.}
\label{Tab:one}
\begin{ruledtabular}
\begin{tabular}{@{}ccccccc@{}}
& LDA & GGA \\\hline\hline
 N\'{e}el   &  -39   &  -113  \\
 double stripe   &  -101  &  -176  \\
 stripe  &  -108  &  -180  \\
 ferromagnetic   &  -89   &  -186  \\\hline
 N\'{e}el(II)   &  -22   &  -88  \\
 double stripe(II)   &  -103  &  -182  \\
 stripe(II)  &  -111  &  -190 \\
 ferromagnetic(II)   &  -64   &  -148
\end{tabular}
\end{ruledtabular}
\end{table}

\begin{figure}[htbp]
\includegraphics[width=0.48\textwidth]{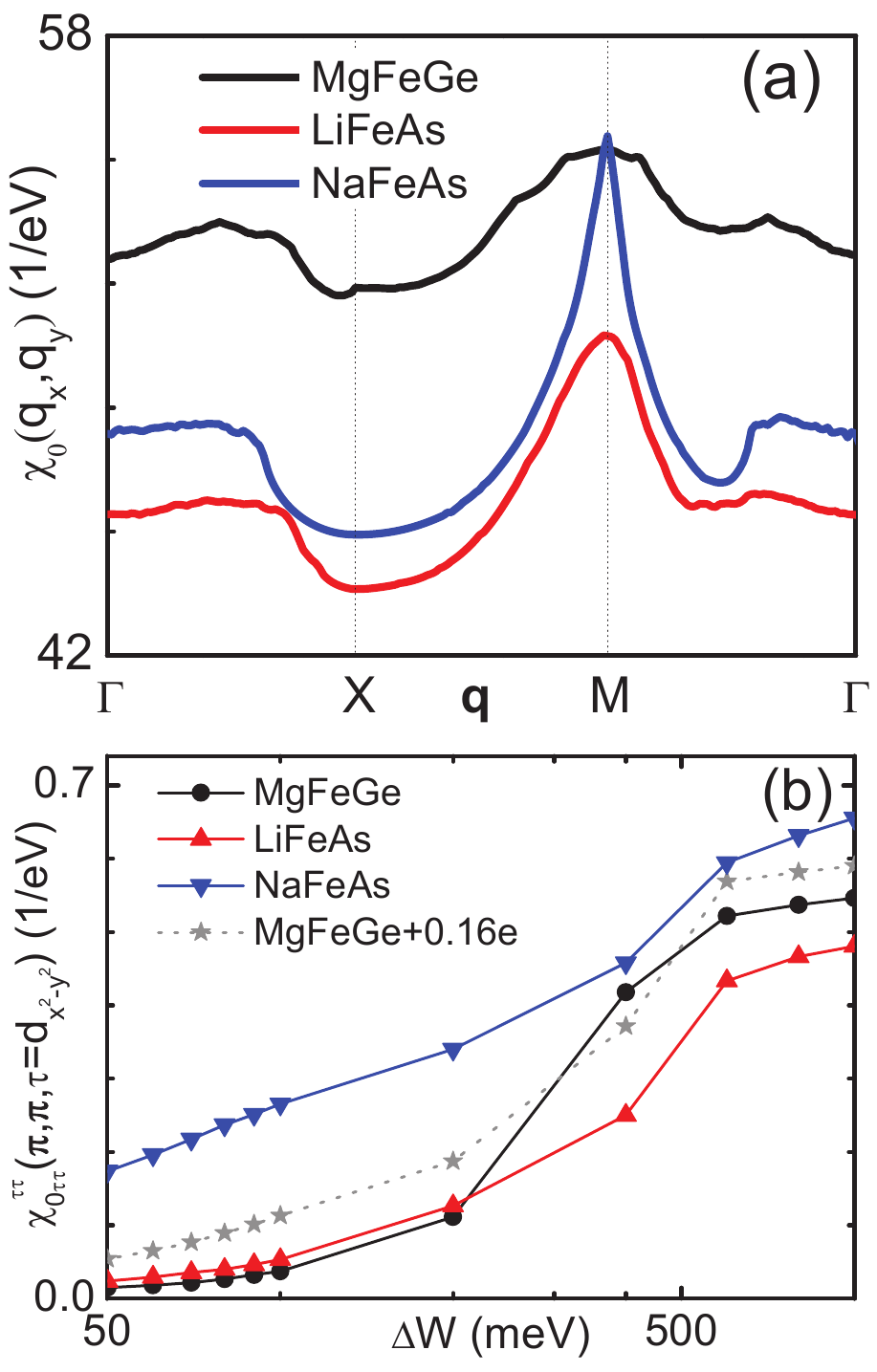}
\caption{(Color online) (a) Pauli susceptibilities within constant matrix elements approximation for MgFeGe, LiFeAs and NaFeAs along the path in the momentum space of $\Gamma(0,0)$-$X(\pi,0)$-$M(\pi,\pi)$-$\Gamma(0,0)$. (b) The intraorbital contributions to the Pauli susceptibilities from d$_{x^2-y^2}$ orbitals at $(\pi,\pi)$ as a function of width of the energy windows chosen around Fermi Level for MgFeGe, LiFeAs, NaFeAs, and MgFeGe with electron doping of 0.16e/Fe. The results are qualitatively the same in $q_z=0$ and $q_z=\pi$ plane.}
\label{Fig:one}
\end{figure}

First, we present comparisons of total energies among different types of magnetically ordered states, such as 1) N\'{e}el state with intralayer N\'{e}el antiferromagnetic and interlayer ferromagnetic ordering, 2) double stripe state with intralayer double-stripe-type antiferromagnetic and interlayer ferromagnetic ordering, 3) stripe state with intralayer stripe-type antiferromagnetic and interlayer ferromagnetic ordering, 4) ferromagnetic state with both interlayer and intralayer ferromagnetic ordering, and 5) N\'{e}el(II) state with intralayer N\'{e}el antiferromagnetic but interlayer antiferromagnetic ordering, 6) double stripe(II) state with intralayer double-stripe-type antiferromagnetic but interlayer antiferromagnetic ordering, 7) stripe(II) state with intralayer stripe-type antiferromagnetic but interlayer antiferromagnetic ordering, 8) ferromagnetic(II) states with intralayer ferromagnetic but interlayer antiferromagnetic ordering.

The results are summarized in Table~\ref{Tab:one}. The first four magnetic states are corresponding to those studied in Ref.~\onlinecite{Jeschke2013}. It is found that the total energies of these four states obtained within GGA are consistent with previous results although full potential linearized augmented plane wave method as implemented in WIEN2k is used here instead of full potential localized orbital method used in Ref.~\onlinecite{Jeschke2013}. Indeed, the ferromagnetic state gives lowest total energy. However, if we compare the total energies among these four states obtained within LDA, the ground state is dramatically changed to stripe state, indicating the strong dependence of the spin state on the functionals. Furthermore, we investigate the effect of interlayer exchange coupling by changing the interlayer spin configurations from ferromagnetic to antiferromagnetic arrangements. It is shown that the stripe(II) state with intralayer stripe-type antiferromagnetic and interlayer antiferromagnetic ordering gives the lowest total energy, regardless of the functionals one chooses. On the other hand, we also calculate the total energies of stripe state and ferromagneitc state based on optimized lattice structures within both GGA and LDA. We find the stripe-type antiferromagnetic state always prevails over the ferromagnetic one. In fact, this result is consistent with the weak coupling theory within constant matrix elements approximation where strong $(\pi,\pi)$ instability is observed in the momentum-dependent Pauli susceptibility (See Fig.~\ref{Fig:one}~(a)) which implies a strong tendency toward stripe-type antiferromagnetic state in MgFeGe. Then, the question remains, why neither magnetization nor superconductivity is observed experimentally? Does it really suggest that MgFeGe do serve as a true counterexample against the itinerant scenario?

\begin{figure*}[htbp]
\includegraphics[width=0.96\textwidth]{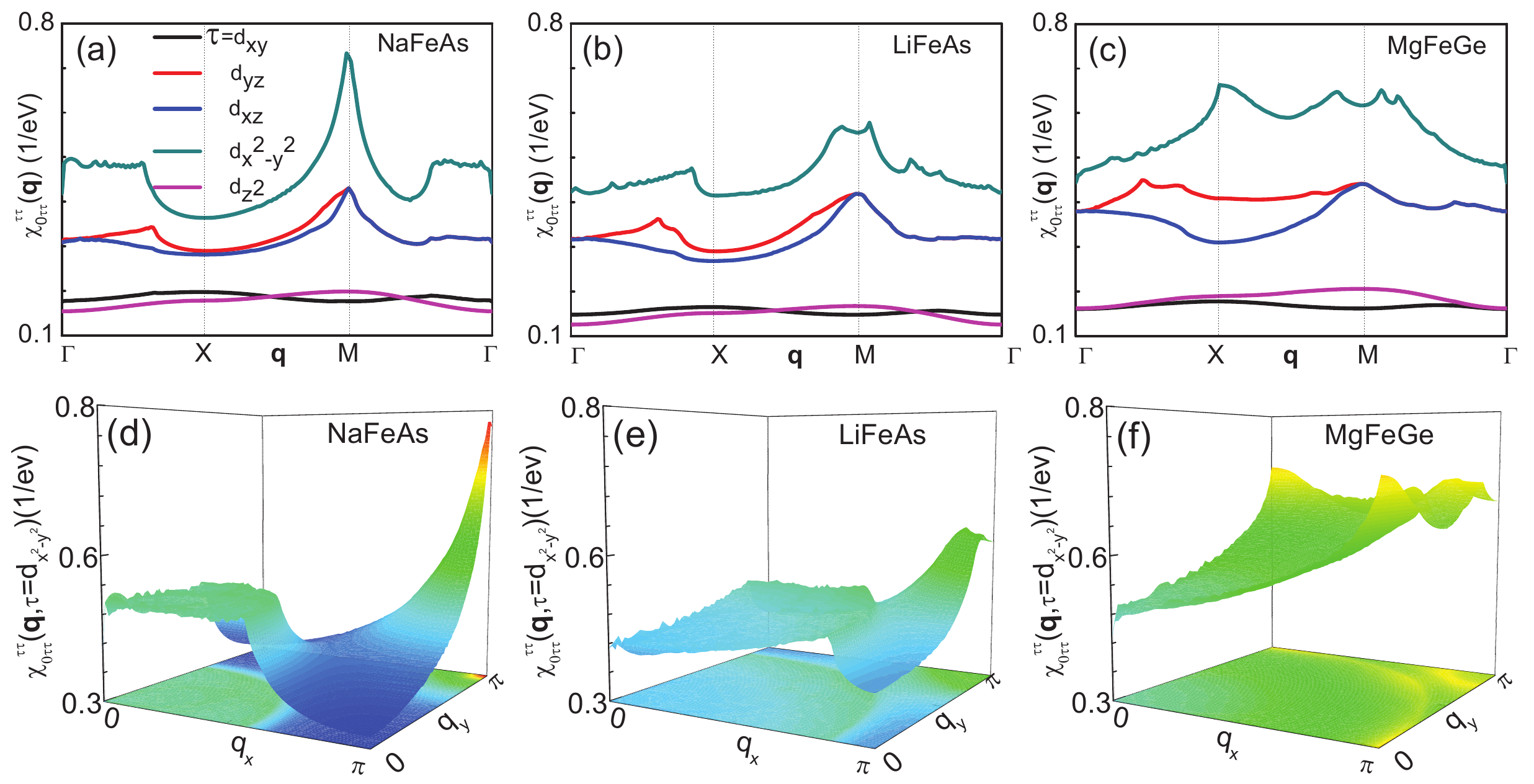}
\caption{(Color online) Five intraorbital contributions to the susceptibilities along the path in the momentum space of $\Gamma(0,0)$-$X(\pi,0)$-$M(\pi,\pi)$-$\Gamma(0,0)$ in NaFeAs (a), LiFeAs (b), and MgFeGe (c). The dominating contributions from d$_{x^2-y^2}$ orbital in NaFeAs (d), LiFeAs (e), and MgFeGe (f) in the $q_x-q_y$ plane. The two-dimensional contour maps are on the bottom.}
\label{Fig:two}
\end{figure*}

\subsection{Theory from itinerant scenario\label{itinerant}}
The above problems can be resolved after orbital contributions to the Pauli susceptibility are properly taken into account. The Pauli susceptibility~\cite{Graser} is defined as
\begin{eqnarray}
  \chi^{pr;st}_{0} (k,\omega)&=& -\frac{1}{N} \sum_{k,\mu\nu}\frac{a^{s}_{\mu}(k)a^{p*}_{\mu}(k)a^{r}_{\nu}(k+q)a^{t*}_{\nu}(k+q)}{\omega+E_{\nu}(k+q)-E_{\mu}(k)+i0^{+}} \nonumber \\
  &&\times [f(E_{\nu}(k+q))-f(E_{\mu}(k))]
\end{eqnarray}
where matrix elements $a^{s}_{\mu}(k)=\langle s|\mu k \rangle$ connect the orbital and the band spaces and are the components of the eigenvectors obtained from diagonalization of an effective tight-binding Hamiltonian derived from the DFT band structure via construction of Wannier orbitals~\cite{wannier1,wannier2}. Here $f(E)$ is the Fermi distribution function, $p,r,s,t$ are the orbital indices and $\mu,\nu$ the band indices.

The crude approximation of constant matrix elements assumes that all the orbitals give the same contributions to each band at every $k$ point by taking all the matrix elements to be $1$, which results in the inconsistencies between the weak coupling theory and the experiments. As can be seen in Fig.~\ref{Fig:one}~(a), all three compounds exhibit prominent instability at $(\pi,\pi)$ and the instability is even stronger in MgFeGe than in LiFeAs.

However, the situation is dramatically changed if the approximation is removed. Since the magnetism and the superconductivity are dominated by the intraorbital scattering~\cite{zhangfete,leeshim,Kuroki}, we only show in Fig.~\ref{Fig:two} the intraorbital contributions to the Pauli susceptibility. Figs.~\ref{Fig:two} (a), (b), and (c) present the orbital-resolved susceptibilities along the path in the momentum space of $(0,0)$-$(\pi,0)$-$(\pi,\pi)$-$(0,0)$ for NaFeAs, LiFeAs, and MgFeGe, respectively. It is found that, on one hand, all three compounds possess a common feature that the susceptibilities of d$_{xy}$ and d$_{z^2}$ orbitals are much smaller than those of d$_{x^2-y^2}$, d$_{xz}$, and d$_{yz}$ orbitals, indicating that the physical properties are mainly controlled by the latter three orbitals. Here $x$, $y$, $z$ are along $a$, $b$, $c$ directions of the original unit cell with two iron atoms, respectively. On the other hand, remarkable differences can be observed among these three compounds. While d$_{x^2-y^2}$ orbital exhibits strong instability at wave vector of $(\pi,\pi)$ in NaFeAs (see also Fig.~\ref{Fig:two} (d)), implying a strong tendency toward the stripe-type antiferromagnetic ordering, the instability of d$_{x^2-y^2}$ orbital in LiFeAs is significantly suppressed (see also Fig.~\ref{Fig:two} (e)). Since appearance of magnetism requires an instability exactly at $(\pi,\pi)$ stronger than a threshold, remaining peaks around $(\pi,\pi)$ in all three orbitals (d$_{x^2-y^2}$, d$_{xz}$, and d$_{yz}$) of LiFeAs (see Fig.~\ref{Fig:two} (b)) indicate a tendency toward superconductivity at sufficiently low temperature. Nevertheless, the behavior of Pauli susceptibility within constant matrix element approximation in NaFeAs and LiFeAs as shown in Fig.~\ref{Fig:one} (a) still qualitatively reflects the fact that the orbitally resolved susceptibilities tell as presented in Figs.~\ref{Fig:two} (a) and (b).

In MgFeGe, drastic differences can be detected between the Pauli susceptibility within constant matrix element approximation and the orbitally resolved susceptibility. While a strong and broadened peak is present at $(\pi,\pi)$ in the Pauli susceptibility within constant matrix element approximation as shown in Fig.~\ref{Fig:one} (a), multiple competing humps appear in all the dominating intraorbital contributions to the susceptibility as shown in Fig.~\ref{Fig:two} (c). In other words, no prominent peak can be found in the susceptibilities of d$_{yz/xz}$ and d$_{x^2-y^2}$ orbitals, compared to those in NaFeAs and LiFeAs (see Figs.~\ref{Fig:two} (a) and (b)). If it would be acceptable that multiple weak humps appearing at momentum vectors of $(\pi,0)$, $(0,\pi)$ and $(\pi,\pi)$ in d$_{x^2-y^2}$ and d$_{xz/yz}$ orbitals can still be viewed as the evidence for the weak and competing tendency towards various magnetic ordered states, it would be still understandable why the total energies of double-stripe-type and stripe-type antiferromagnetic states are quite close in both spin polarized GGA and LSDA calculations (see Table~\ref{Tab:one}) and why those states give second lowest and lowest total energies, respectively, in LSDA calculations. This is due to the fact that the dynamical fluctuations are completely ignored in the DFT calculations. If the dynamical fluctuations among various quantum states are switched on, any instabilities with comparable strength at different momentum vectors as present in the orbitally resolved susceptibilities in MgFeGe can not win the competition against the others at low temperature, which should be the reason from weak coupling itinerant limit why MgFeGe remains nonmagnetic at low temperature.

In Fig~\ref{Fig:one} (b), we present the intra-orbital Pauli susceptibilities of d$_{x^2-y^2}$ at $(\pi,\pi)$ as a function of width of the energy window chosen around Fermi level. It is found that the susceptibility of MgFeGe calculated within large energy window is larger than that of LiFeAs, while it surprisingly becomes smaller as the energy window is shrunk. Since the Pauli susceptibility within small energy window chosen in the vicinity of the Fermi level reflects the nesting properties of the Fermi surface, our results strongly indicate that MgFeGe is not a superconductor due to the poorer Fermi surface nesting than LiFeAs. Furthermore, we find that the Pauli susceptibilities of d$_{xz/yz}$ orbitals at $(\pi,\pi)$ remain almost the same between LiFeAs and MgFeGe within small energy window, implying that d$_{x^2-y^2}$ orbital dominates the contrast low temperature behaviors of MgFeGe, LiFeAs, and NaFeAs.

\begin{figure}[htbp]
\includegraphics[width=0.48\textwidth]{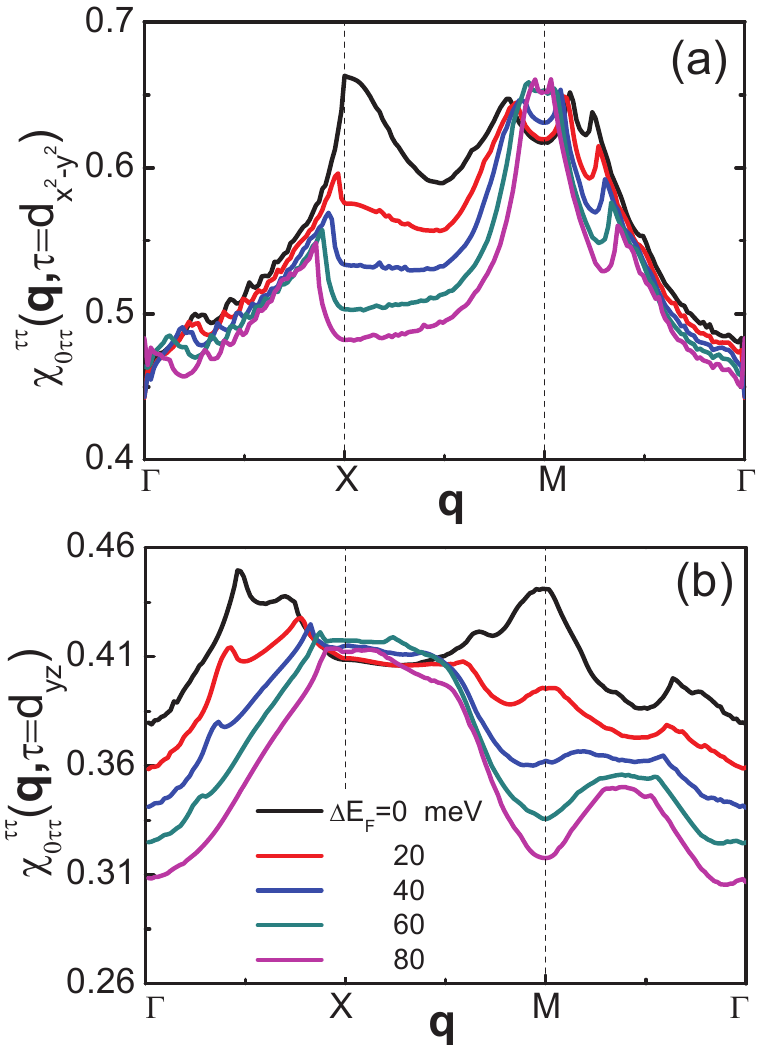}
\caption{(Color online) Evolution of intraorbital contributions to the susceptibility at different shifted Fermi energies. (a) d$_{x^2-y^2}$ and (b) d$_{yz}$ orbital.
$\Delta E_F = 0, 20, 40, 60, 80$~meV, corresponding to the electron doping of $0, 0.05, 0.11, 0.16, 0.21$ per Fe atom. The path in the momentum space is $\Gamma(0,0)$-$X(\pi,0)$-$M(\pi,\pi)$-$\Gamma(0,0)$.}
\label{Fig:three}
\end{figure}

\subsection{Predictions\label{Predictions}}
From above comparisons between different compounds and the analyses on the orbitally resolved susceptibility, we learn that a pronounce instability around $(\pi,\pi)$ in the orbitally resolved susceptibility is the precursor for the superconducting state in MgFeGe. Therefore, in order to make MgFeGe a possible new iron-based superconductor, we try different ways to increase the instability around $(\pi,\pi)$ in MgFeGe, such as shift the Fermi level up and down which can be viewed as electron and hole doping, respectively, as well as move the Ge atom close to or away from Fe plane which can be viewed as applying chemical or external pressure. We find that lifting the Fermi level up, served as electron doping, is an efficient way to enhance the instability around $(\pi,\pi)$ in d$_{x^2-y^2}$ orbital. As shown in Fig~\ref{Fig:three}, we shift the Fermi level up by 20, 40, 60, and 80~meV, which is corresponding to dope 0.05, 0.11, 0.16, 0.21~electrons per Fe. It is found that in both d$_{x^2-y^2}$ and d$_{yz}$ orbitals, multiple-peak structures with peaks of comparable strength appearing in the undoped case (see the black line in Fig~\ref{Fig:three}~(a) and (b)) vanishes. Instead, prominent peaks around $(\pi,\pi)$ and $(\pi,0)$ occur in d$_{x^2-y^2}$ and d$_{yz}$ orbitals, respectively, due to the suppression of other peaks as electron doping increases. Nevertheless, the peak around $(\pi,0)$ in d$_{yz}$ is weaker and more broadened than that around $(\pi,\pi)$ in d$_{x^2-y^2}$, inferring that the instability around $(\pi,\pi)$ may play a dominating role in determining the low temperature physics and consequently lead to a possible superconducting state as the Fermi surface nesting becomes better (See Fig~\ref{Fig:one} (b)).

\section{Discussions\label{Discussions}}
In fact, preliminary effort has been made in looking for a superconducting state in MgFeGe. However, it is failed after electron doping by cobalt and hole doping in the form of Mg$_{1-x}$FeGe~\cite{hosonomgfege}. In our calculations, hole doping will significantly enhance the ferromagnetic instability in d$_{xz/yz}$ orbital but slightly enhance the stripe-type antiferromagnetic instability in d$_{x^2-y^2}$ orbital, resulting in the failure of hole doping. On the other hand, it is known from a first principles investigation that Co substitution of Fe will impose disorder effect on the parent compound and does not merely change the carrier density~\cite{Berlijn}. Thus, we propose based on our above investigation that it would be an efficient way to do an electron doping by increasing the content of Mg or in the form of Mg$_{1-x}$Al$_x$FeGe, Mg$_{1-x}$La$_x$FeGe, etc.

Recently, an LDA+DMFT method has been applied to study the spin dynamics in MgFeGe~\cite{zpyin}. It is found that spin fluctuations are ferromagnetic. However, first, it is hard to believe that a local and $q$-independent correction due to the electronic interactions will generate dramatically an instability at $(0,0)$ which doesn't exist in zero order susceptibility while suppress all the others. Second, the Hubbard $U$ interaction used in the LDA+DMFT study is 5~eV which is too large compared to the $U$ values derived from a constrained random phase approximation method for LiFeAs~\cite{Miyake2010}. Moreover, our constrained LDA calculations show that the local interaction $U$ is slightly smaller in MgFeGe than in LiFeAs. As is well-known from the mean-field phase diagram of Hubbard model, large $U$ favors the ferromagnetic solution~\cite{hirsch,Jiexu}. Third, it is always a problem in determining precisely the Fermi level or the filling in LDA+DMFT calculations. As we know from our orbitally resolved susceptibility calculations, slightly shifting downward the Fermi level will enhance the ferromagnetic instability.

\section{Conclusions\label{Conclusions}}
1)~We show that the contrasting low-temperature behaviors among MgFeGe, LiFeAs, and NaFeAs are dominated by the distinct structures of the intraorbital contributions to the particle-hole scattering from d$_{x^2-y^2}$ orbital. This implies that an effective single orbital model may be sufficient to capture physics of the 111 family of iron-based superconductors which possess complicated band structures compared to those of high-T$_c$ cuprates. Moreover, competing instabilities coexisting in the same orbitals which can be effectively viewed as the magnetic frustrations from itinerant electrons may also be applicable to understand other iron-based superconductors with nonmagnetic parent states. 2)~Our results resolve the existing confusions in 111 family of iron-based superconductors and indicate that the physics of iron-based superconductors can be understood from weak coupling itinerant limit where electronic correlations play subsidiary roles. 3)~We propose a possible way to synthesize a new iron-based superconductor based on the calculations of electron doping to MgFeGe. As our results are robust irrespective of the functionals one chooses, in contrast to the previous consensus that LDA and GGA give inconsistent answers~\cite{Mazinsingh}, we propose that investigating the intraorbital particle-hole scattering can capture the overall trends in the same family of iron-based superconductors and should be a promising way for understanding existing or predicting new iron-based superconductors.

\section{Acknowledgements\label{Acknowledgements}}
We thank Dr. Harald O. Jeschke and Prof. Fan Yang for discussions. This work is supported by National Natural Science Foundation of China (No. 11174219), Program for New Century Excellent Talents in University (NCET-13-0428), Research Fund for the Doctoral Program of Higher Education of China (No. 20110072110044) and the Program for Professor of Special Appointment (Eastern Scholar) at Shanghai Institutions of Higher Learning as well as the Scientific Research Foundation for the Returned Overseas Chinese Scholars, State Education Ministry.


\begin{thebibliography}{99}
\bibitem{ourresult} We use experimental crystal structures of MgFeGe~\cite{hosonomgfege}, LiFeAs~\cite{Tapplifeas} and NaFeAs~\cite{dainafeas}. The full potential linearized augmented plane wave method as implemented in wien2k~\cite{balaha} is employed to calculate the electronic structures. The tetrahedron method is used and a mesh with total 40000 k points is sampled for Brillouin-zone integration in both the nonmagnetic and different spin configuration calculations.
\end{thebibliography}
\end{document}